\documentclass[twocolumn,showpacs,preprintnumbers,amsmath,amssymb,prb,aps]{revtex4-1}
\usepackage{epsfig}
\usepackage{epstopdf}
\usepackage{graphicx}
\usepackage{dcolumn}
\usepackage{bm}
\usepackage{slashbox}
\usepackage{textcomp}
\usepackage{array}
\usepackage{subcaption}
\usepackage{booktabs}

\begin{document}

\title{Exploring the best scenario for understanding the high temperature thermoelectric behaviour of Fe$_{2}$VAl}
\author{Shamim Sk$^{1,}$}
\altaffiliation{Electronic mail: shamimsk20@gmail.com}
\author{P. Devi$^{2}$}
\author{Sanjay Singh$^{2,3}$}
\author{Sudhir K Pandey$^{4}$}
\affiliation{$^{1}$School of Basic Sciences, Indian Institute of Technology Mandi, Kamand - 175005, India}
\affiliation{$^{2}$Max Planck Institute for Chemical Physics of Solids, N\"{o}thnitzer Str. 40, 01187 Dresden, Germany}
\affiliation{$^{3}$School of Materials Science and Technology, Indian Institute of Technology (Banaras Hindu University), Varanasi-221005, India}
\affiliation{$^{4}$School of Engineering, Indian Institute of Technology Mandi, Kamand - 175005, India}

\date{\today}

\begin{abstract}
Heusler-type Fe$_{2}$VAl compound is a promising thermoelectric candidate with non-magnetic ground state. The present work investigates the Seebeck coefficient (S) of Fe$_{2}$VAl in the temperature region 300 to 620 K with the help of experimental and theoretical tools. The experimental value of S is observed $\sim$ -130 $\mu$V/K at 300 K. Afterthat, the magnitude of S decreases gradually as the temperature increases. At T = 620 K, the value of S is found to be $\sim$ -26 $\mu$V/K. In order to understand the behaviour of the experimentally observed S value, the band-structure and density of states calculations are performed by using LDA, PBE, PBEsol, mBJ and SCAN within density functional theory. All the above mentioned exchange-correlation (XC) functionals (except mBJ) predict the semi-metal like behaviour of the compound, whereas the mBJ gives the indirect band gap of $\sim$ 0.22 eV having the well agreement with experimentally observed value. The temperature dependence of S for Fe$_{2}$VAl is also calculated with the help of all the five mentioned functionals individually. The best XC functional is investigated for searching the new thermoelectric materials by taking Fe$_{2}$VAl as a case example through this study. The best matching between experimental and calculated values of S as a function of temperature is observed by setting the mBJ band gap with the band-structure of PBEsol or SCAN. Therefore, the present study suggests that the band-structure of PBEsol or SCAN with mBJ band gap can be used for searching the new thermoelectric materials.

\end{abstract}

\maketitle

\section{Introduction} 
 Today's world is seeking for an alternative sources of energy due to the fast utilization of natural energy sources. At the present time, energy demands has been significantly increasing. But, the natural source of energy is limited and hence it is a time to think for an alternative energy sources. There are various energy generator, such as solar cells\cite{solarcell}, bio-mass, hydro-power, nuclear power plants\cite{nuclearpower}, thermoelectric generator (TEG)\cite{teg1,teg2} have been developed in the last few decades. Nowadays, almost 90$ \% $  of the total power supply still depends on fossil fuels.\cite{fossilfuel} It is important to note here, more than 50$ \% $  of the energy expenditure of fossil fuels is released to the atmosphere in the form of waste heat by engines. This waste heat can be converted into useful electricity by the means of TEG, which is based on thermoelectric (TE) materials. The presently available TEGs has their low efficiency\cite{progress} and they are strongly dependent on performance of TE materials. The efficiency of the TE materials is defined as a dimensionless parameter, figure-of-merit (ZT)\cite{zt1,zt2} 
\begin{equation}
ZT = \frac{S^{2}\sigma T}{\kappa},
\end{equation}
where S is the Seebeck coefficient, $ \sigma $ is the electrical conductivity, T is the absolute temperature and $ \kappa(=\kappa_{e} + \kappa_{ph}) $ is the thermal conductivity in which the electronic ($\kappa_{e}$) and phononic ($\kappa_{ph}$) part is involved. The requirement of high TE efficiency is high ZT value of the materials, which makes the TEG competitive with other alternative energy sources. The high ZT materials should have high power facror ($ S^{2}\sigma $) with low $ \kappa $ value.\cite{highzt} But, S, $ \sigma $ and $ \kappa_{e} $ are strongly correlated through the carrier concentration. \cite{coupled} Therefore, the optimization of S, $ \sigma $ and $ \kappa $ to get a high ZT value is challenging task for today's researchers. The square dependent of S in equation (1) makes it is an important quantity to enhance the materials more efficient. Simply, S is defined as the ratio of the induced voltage difference in responce to a temperature gradient (dV/dT) between two points within the material. It is also known as thermopower of the thermoelctric materials.

Nowadays, there are very less TE materials having high ZT value.\cite{bite,pbte,sige} The traditionally available high ZT thermoelectric materials consist of Bi-Te\cite{bite}, Pb-Te\cite{pbte} and Si-Ge\cite{sige} compound, mainly. But, Bi, Te and Pb are easily decomposed and oxidised at higher temperature. Also, difficulty comes from some of these elements (e.g. Pb, Te), because of toxic in nature, rare and costly. Therefore, finding out the alternative materials with high ZT value is required.

There are two ways for searching the new high ZT thermoelectric materials. These are experimental and computational path. Experimentally, The ZT can be improved by optimizing the existing TE materials or searching the new materials, but this process is time consuming and costly. On the other hand, the new materials for high ZT can be searched computationally, by changing the parameter in the calculation in such a way that the results should be reliable. This method is less time consuming as well as low cost. In computational method, the first-principle density functional theory (DFT) found as the most powerfull tool for electronic structure calculation. In this theory, the Kohn-Sham (KS) equation is solved self-consistently by considering single electron wave functions.\cite{kohn} The electron-electron interaction part in KS equation is approximated as exchange-correlation (XC) functionals. Many XC functionals have developed for better approximation in last few decades with their own merit and demerit. At present time, The most useful XC functionals are Local density approximation of Perdew and Wang-1992 (LDA-PW92)\cite{lda92}, generalized gradient approximation of Perdew-Burke-Ernzerhof (GGA-PBE)\cite{pbe} and newly developed PBEsol\cite{pbesol}, mBJ\cite{mbj} and SCAN\cite{scan} in condensed matter physics. It is seen from the literature, different functionals give the different features of the band-structure.\cite{shastri} It is also important to  mention that, mBJ gives the more correct value of the band gap\cite{tran}, whereas all the others four functionals underestimate the band gap of the materials significantly. The researchers who are working on the field of thermoelctric materials, they considered any one of the functionals as a standard for studying the thermoelectric properties of the materials.  For instance, Shastri \textit{et al.}\cite{shastri} have qualitatively studied the thermoelctric properties of the materials by using all the above maintained functionals. Sharma \textit{et al.}\cite{sharma} have worked on TE properties of full-Heusler alloy by using PBEsol functional. D. do \textit{et al.}\cite{ddo} have reported the TE properties of full-Heusler compounds by using GGA/PBE functional. Rai \textit{et al.}\cite{rai} have used mBJ potential for describing the thermoelectric properties of Fe$_{2}$VAl.  

As per our knowledge from literature survey, there are no such benchmark of XC functionals to study the thermoelectric properties of the materials. Therefore, quantitative description for searching the new high ZT materials by using the available XC functionals is required. As we know that, the different functionals give the different features of the dispersion curve and this is implicitly or explicitly affected the thermoelectric parameters, i.e $\sigma$, $\kappa_{e}$ and S. The electrical conductivity of the materials can be expressed by the following equation\cite{ashcroft},
\begin{equation}
\boldsymbol{\sigma} = \sum_{n}\boldsymbol{\sigma}^{(n)}
\end{equation}
where,
\begin{equation}
\boldsymbol{\sigma}^{(n)} = e^{2}\int\frac{d\textbf{k}}{4\pi^{3}}\tau_{n}(\varepsilon_{n}(\textbf{k}))\textbf{v}_{n}(\textbf{k})\textbf{v}_{n}(\textbf{k})(-\frac{\partial f}{\partial \varepsilon})_{\varepsilon=\varepsilon_{n}(\textbf{k})}
\end{equation}

where, $\textbf{v}_{n}(\textbf{k})$ is the mean velocity of an electron in a level specified by band index n and wave vector \textbf{k} and it is written as\cite{ashcroft},
\begin{equation}
\textbf{v}_{n}(\textbf{k}) = \frac{1}{\hbar}\boldsymbol{\nabla}_{\textbf{k}}\varepsilon_{n}(\textbf{k})
\end{equation} 
where, $\varepsilon_{n}(\textit{\textbf{k}})$ is known as energy band. Therefore, by combining equations (2), (3) and (4), it is evident that the electrical conductivity depends on band-structure of the materials. Similarly, the S and the electronic part of thermal conductivity ($\kappa_{e}$) are also connected with electrical conductivity\cite{ashcroft}. Hence S and $\kappa_{e}$ depend on band-features of the materials explicitly. The above discussion suggests that the different results in TE properties is expected by using different functionals, as they give different E vs k plots of the materials. Hence, the prediction of TE properties by using any one of the XC functional may not give the most accurate results. Thus, the systematic investigation in this direction is required. The motivation of this work is to find out the best scenario of XC functional for studying the TE properties of the materials.

In this work, we have studied the temperature dependence of S for Fe$ _{2} $VAl compound in the temperature region 320 to 620 K experimentally. The elctronic structure calculations are performed by using five XC functionals (LDA, PBE, PBEsol, mBJ and SCAN) within density functional theory. The mBJ gives the indirect band gap of $\sim$ 0.22 eV with the good experimental agreement, whereas all the other functionals predict the semi-metal like behaviour of the compound. We also calculate the temperature dependence of S and then systematically investigate the best functional to study the S vs T by comparing the calculated S with experiment. The LDA, PBE, PBEsol and SCAN give S $\approx$ 3 $-$ 5 $\mu$V/K at 300 K, which is negligibly small as compared to the experimental value of $\sim$ -130 $\mu$V/K, while mBJ gives S $\approx$ 25 $\mu$V/K at 620 K, in quite good agreement with experiment ($\sim$ -26 $\mu$V/K). The best matching between experimental and calculated value of S as a function of temperature is observed when mBJ band gap is used with the band-features (E vs k plots) of PBEsol or SCAN.

\section{Experimental and computational details}
Polycrystalline ingots of Fe$_{2}$VAl was prepared by melting appropriate quantities of Fe, V, and Al of 99.99\,\% purity in an arc furnace. The prepared ingots were then annealed at 1073~K for 5 days to obtain homogeneity and subsequently quenched into ice water. The room temperature x-ray diffraction confirms the Heusler L2$_1$ structure of sample with lattice parameters 5.76 \AA, which is in well agreement with the literature.\cite{boscow81} The S of Fe$ _{2} $VAl compound was measured in the temperature region 300 $-$ 620 K by using the home-made experimental setup.\cite{ashutosh} For the measurement, the sample with pellet form was used with thickness of 3.9 mm and cross-sectional area of 14.2 mm$ ^{2} $.

To examine the experimentally obtained value of S, we explore the electronic and transport properties of the compound. The electronic properties are studied with density functional theory\cite{kohn1,kohn2} using the full-potential linearized augmented plane wave method (FP-LAPW) as implemented in WIEN2k\cite{wein2k} code. The five functionals are used as an exchange-correlation part for the self-consistent calculations. These are LDA of Perdew-Wang-1992 (LDA)\cite{lda92}, GGA of Perdew-Burke-Ernzerhof (PBE)\cite{pbe} and recently developed PBEsol\cite{pbesol}, mBJ\cite{mbj} and  SCAN.\cite{scan} The LDA is used as a correlation part with modified Becke-Johnson (mBJ) potential. The different values of optimized lattice parameter for different functionals (which can be found in ref.\cite{shastri}) are used as an initial input value for the ground-state electronic calculations. 

Fe$ _{2} $VAl has the structure of  \textit{L}2$ _{1} $ with spacegroup \textit{Fm}3\textit{m}. The primitive cell of this compound contains four atoms to form a lattice with the Wyckoff positions Fe1 $(\frac{1}{4},\frac{1}{4},\frac{1}{4})$, Fe2 $(\frac{3}{4},\frac{3}{4},\frac{3}{4})$, V $(0,0,0)$ and Al $(\frac{1}{2},\frac{1}{2},\frac{1}{2})$. The muffin-tin sphere radii R$ _{MT} $ for Fe, V and Al were set to be 2.36, 2.24 and 2.13 bohr, respectively. A k-mesh grid of size 50 $ \times $ 50 $ \times $ 50 was used for the calculations. The self-consistency calculations were continued until the converging criteria is satishfied and in the present case the total charge/cell convergence criteria was set to be less than 0.1 mRy. The temperature dependence of S was studied by using BoltzTraP package, \cite{boltztrap} which is based on constant relaxation time approximation.

\section{Results and Discussion}

First of all, the experimental result of Seebeck coefficient (S) for Fe$ _{2} $VAl compound is discussed. Fig. 1 shows the temperature dependence of S in the temperature region 300 to 620 K. At T = 300 K, the observed value of S is $\sim$ -130 $ \mu $V/K. Afterthat, the magnitude of S decrese as the temperature increases and at 620 K the value of S observed is $\sim$ -26 $ \mu $V/K. The similar type of behaviour is studied by K. Kurosaki \textit{et al}. \cite{kurosaki} For stoichiometric Fe$ _{2} $VAl, the positive values of S are reported by many groups\cite{positives1, positives2, positives3, positives4} and also the negative values of S with low magnitude ($\sim$ -20 to $\sim$ -10 $\mu$V/K in the temperature range 300 to 600 K ) is also reported by Mikami \textit{et al}\cite{mikami}  Our sample shows the negative values of S with higher magnitude through the temperature range studied, which indicates the slight off-stoichiometry of the sample. The negative values of S indicate the n-type behaviour of the sample.  

\begin{figure}
 
\includegraphics[width=0.90\linewidth, height=7.0cm]{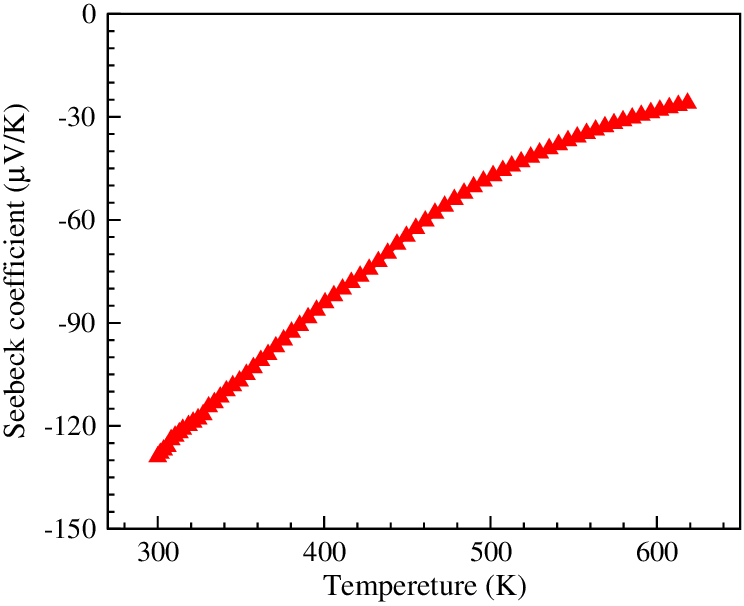} 
\caption{\small{Temperature dependence of Seebeck coefficient in Fe$_{2}$VAl.}}
\end{figure}

\begin{figure*} 

\begin{subfigure}{0.4\textwidth}
\includegraphics[width=0.985\linewidth, height=6.0cm]{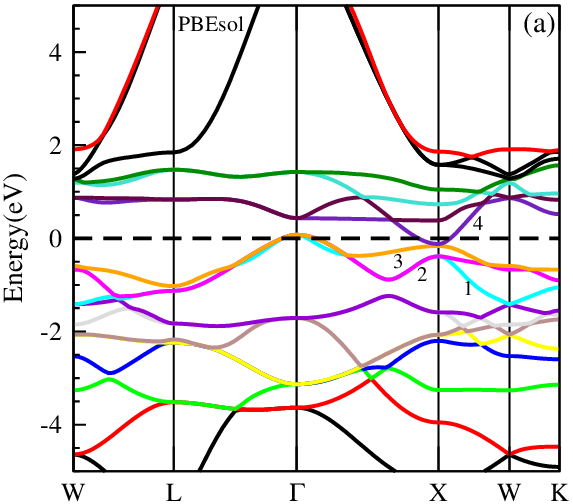} 
\end{subfigure}
\begin{subfigure}{0.4\textwidth}
\includegraphics[width=0.98\linewidth, height=6.0cm]{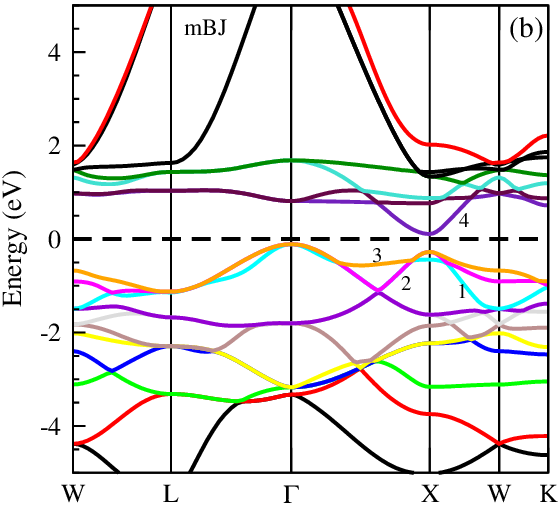}
\end{subfigure} 
\caption{\small{The calculated band-structure of Fe$_{2}$VAl obtained from (a) PBEsol and (b) mBJ.}}
\label{fig:image2}
\end{figure*}

To understand the behaviour of the experimentally measured S, we have carried out the electronic structure calculations. The dispersion curve and density of states (DOS) are computed by using five different exchange-correlation (XC) functionals. The LDA, PBE, PBEsol and SCAN give the similar results in electronic calculation, whereas different result is observed in case of mBJ. Therefore, the results of PBEsol and mBJ calculations are described below. 

Fig. 2 (a) and (b) show the calculated dispersion curve obtained using PBEsol and mBJ, respectively. It is clear that from the dispersion curve of PBEsol, the bottom of the conduction band (CB) is just below the E$ _{F} $ at the X-point and which has lower energy than top of the valence band (VB). This predicts the semi-metal like nature of the compound. The direct gap between top of the VB and second bottom of CB observed is $\sim$ 0.36 eV. Fig. 2(b) exhibits the dispersion curve calculated from mBJ potential. The mBJ is known to give the more accurate band gap of the semiconductors and insulators.\cite{tran}  In the figure, the top of the VB and the bottom of the CB occur at $\Gamma$ and X-point, respectively. The observed value of indirect band gap is $\sim$ 0.22 eV, which gives the well agreement with the experimental value.\cite{okamura} We set the zero of chemical potential at middle of the band gap, which is represented as dashed line. The mBJ features of the band-structure is different from that of PBEsol features. The four bands (represented as 1, 2, 3 and 4) near the Fermi level are mainly contributed to the S for both the functionals are expected in the tempertaure range studied. Here, the top of the VB is triply degenerated (bands: 1, 2 and 3) at $\Gamma$-point for both the functionals. In case of PBEsol, at X-point, the top of the VB (band 3) is non-degenerated and this band also touches the bottom of the CB. But, in case of mBJ, the top of the VB band (at X-point) is doubly degenerated (bands: 2 and 3) with giving the finite gap from bottom of the CB. Therefore, even we give the same band gap, this changes in features of band-structure near the Fermi level are expecting to give the different S values by using both the functionals. 

\begin{figure*} 
\includegraphics[width=0.96\linewidth, height=10.5cm]{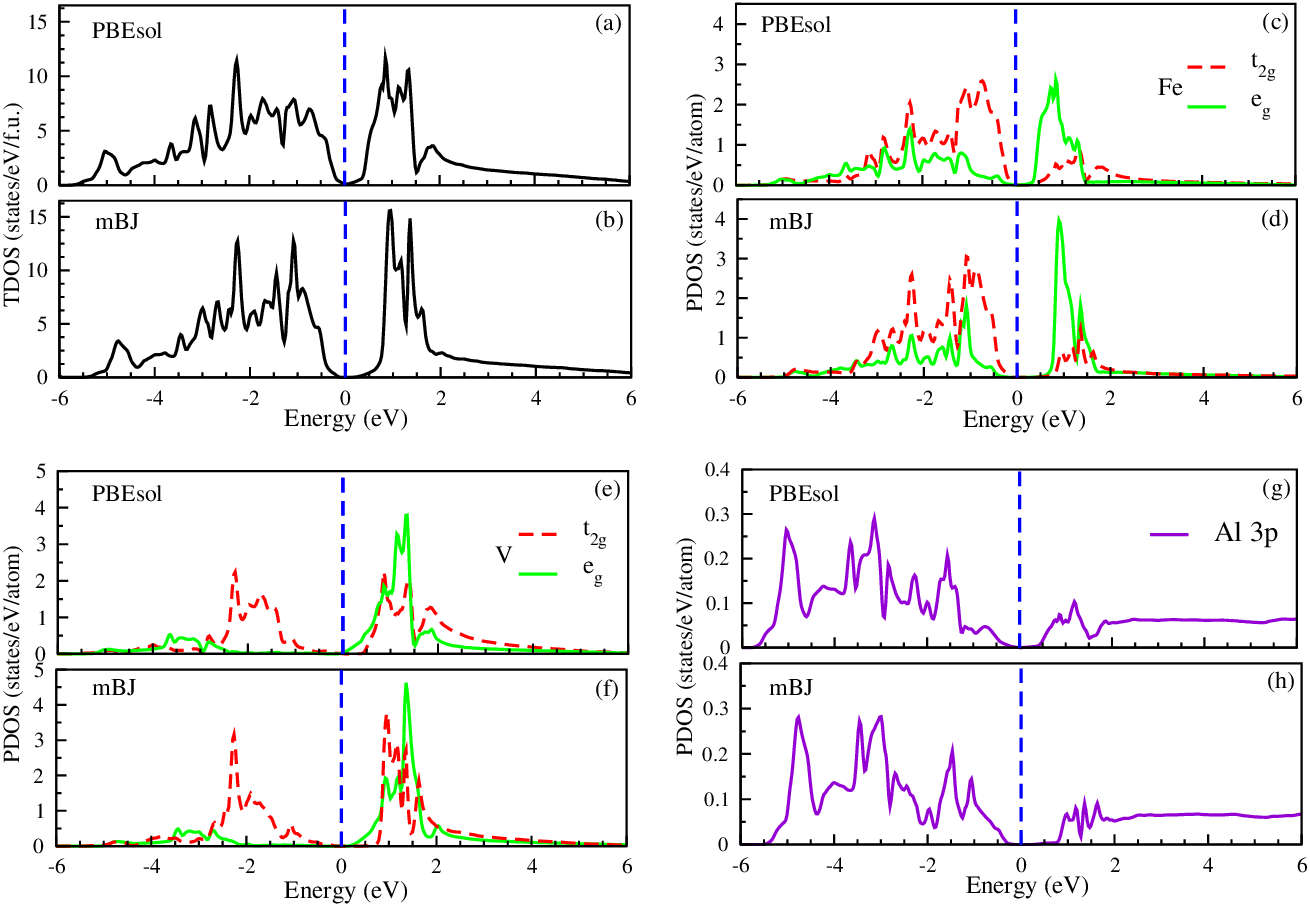} 
\caption{\small{Total and partial density of states plots for Fe$_{2}$VAl obtained using PBEsol and mBJ: (a) $\&$ (b) TDOS, (c) $\&$ (d) PDOS of Fe atom (\textit{3d} orbitals), (e) $\&$ (f) PDOS of V atom (\textit{3d} orbitals), (g) $\&$ (h) PDOS of Al atom (\textit{3p} orbitals). }}
\end{figure*}

Total density of states (TDOS) and partial density of states (PDOS) are also calculated for  Fe$ _{2} $VAl. Fig. 3(a) and (b) exhibit the calculated TDOS by using PBEsol and mBJ, respectively. PBEsol yields the semi-metal like behaviour near the Fermi level. The band gap is clearly seen in case of mBJ as predicted by band-structure calculation. In the low lying energy range ($\sim$ -0.5 to 0 eV) in the VB, the contribution for DOS using PBEsol is more than that of mBJ, which is expected to give the different S values, even we use the same band gap.    

In order to know the contribution of charge carriers to S from different atomic orbitals, we have calculated the PDOS for all the constituent atoms (Fe, V and Al) of Fe$ _{2} $VAl using PBEsol and mBJ potentials, which is shown in fig. 3(c)$-$(h). From the calculation, it is seen that the electronic contribution near the Fermi level comes from Fe \textit{3d} and V \textit{3d} orbitals mainly with negligibly small contribution from Al \textit{3p} orbital. For the simple cubic (SC) lattice, five degenerate \textit{d} orbitals split into three-fold degenerate states \textit{t$ _{2g} $} (\textit{d$ _{xy} $ , d$ _{yz} $ , d$ _{zx} $}) and two-fold degenerate states \textit{e$ _{g} $} (\textit{d$ _{x^{2} - y^{2}} $, d$ _{3z^{2} - r^{2}} $}). Fig. 3(c) and (d) show the contribution of Fe \textit{3d} orbitals near the Fermi level. From the figure, it is clearly observed that in VB the contributions in DOS are mainly from \textit{t$ _{2g} $} states ($\sim$ 70 $\%$) with quite small contribution from \textit{e$ _{g} $} states ($\sim$ 30 $\%$) , whereas in conduction band the contribution from \textit{t$ _{2g} $} and \textit{e$ _{g} $} states are $\sim$ 35 $\%$  and $\sim$ 65 $\%$ , respectively. Fig. 3(e) and (f) exhibit the contribution of V \textit{3d} orbitals in the DOS. It is clear from the figure, in VB the dominant contribution in DOS is observed from \textit{t$ _{2g} $} ($\sim$ 80 $\%$ ) states mainly, whereas in CB the \textit{t$ _{2g} $} and \textit{e$ _{g} $} states are equally contributed in the DOS. For both Fe and V, the hole contribution mainly comes from \textit{t$ _{2g} $} states in VB ($\sim$ -1 eV to 0 eV), whereas the electron contribution comes from \textit{e$ _{g} $} states in CB (0 eV to $\sim$ 1 eV). The dominant contributing states in VB ($\sim$ -1 to 0 eV) and CB (0 to $\sim$ 1 eV) are \textit{t$ _{2g} $} and \textit{e$_{g}$}, respectively. Fig. 3(g) and (h) represent the PDOS plots of Al \textit{3p} given by using PBEsol and mBJ potential. It is seen from the figure the contribution of Al atom in DOS is very less in compare to Fe and V atoms near the Fermi level. Here, it is also important to note that the features and contributions of PDOS near the Fermi level are different for PBEsol and mBJ. Therefore, from the above discussions, it is expected to give the different S values for both the functionals, even we give the same band gap. 

The temperature dependence of S is calculated by using five mentioned XC functionals through the present study. Fig. 4 shows the calculated values of S (using LDA, PBE, PBEsol and SCAN functionals) as a function of temperature. The calculated value of S is found to be 3 $-$ 5 $\mu$V/K at 300 K, which is far away from the experimentally reported value of $\sim$ 40 $\mu$V/K around the room temperature.\cite{positives1,positives2,positives3} When we consider small amount of off-stoichiometry of the compound through the $\mu$ shift upto $\sim$ 20 meV, the value of S is found to be $\sim$ -0.5 to $\sim$ -10 $\mu$V/K at 300 K, which is negligibly small as compare to our experimentally observed value of $\sim$ -130 $\mu$V/K at 300 K. This result suggests the failure of these four functionals in estimating the correct value for the S. Therefore, care should be taken to use these four functionals for searching the new TE materials. 

\begin{figure}
\includegraphics[width=0.91\linewidth, height=7.2cm]{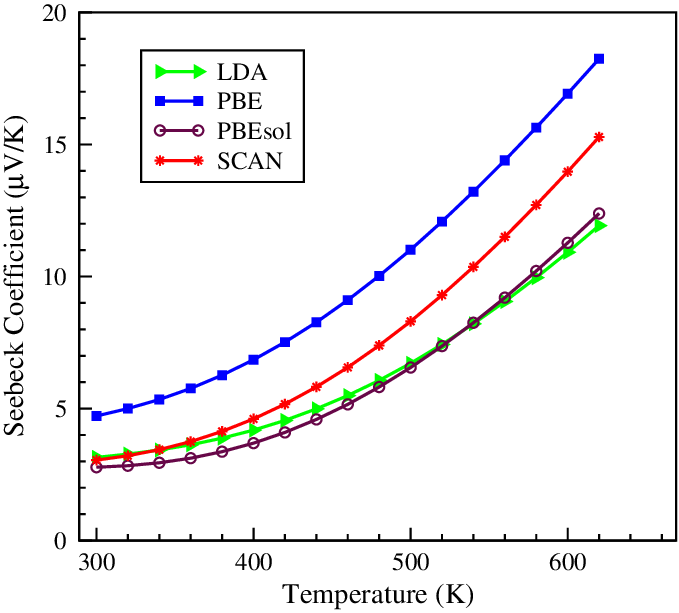} 
\caption{\small{Calculated values of S as a function of temperature obtained using LDA, PBE, PBEsol and SCAN.}}
\end{figure}

\begin{figure}
\includegraphics[width=0.95\linewidth, height=7.15cm]{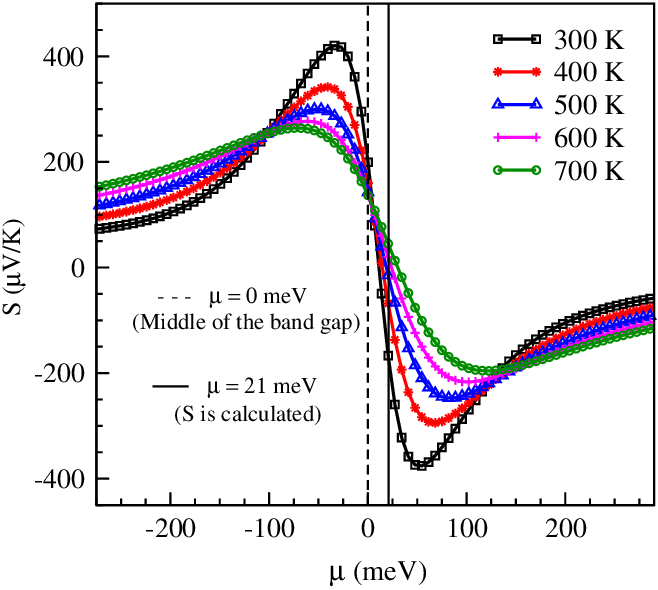} 
\caption{\small{Change in Seebeck coefficient with chemical potential at different temperature obtained using mBJ.}}
\end{figure}

Temperature dependence of S is also calculated using mBJ potential for Fe$_{2}$VAl. The change in S with chemical potential at various temperature is shown in Fig. 5. From the figure, it is evident that if we consider $ \mu $ = 0 meV, which represents the stoichiometry of the compound, the calculated value of S is found to be $\sim$ 200 $\mu$V/K at 300 K, which is very much far away from the experimental reported data\cite{positives1,positives2,positives3}. Then we find the $\mu$ at 300 K for calcuting the S to get a best matching with our experimentally observed value. We found that at $\mu$ $\approx$ 20 meV, which corresponds to small off-stoichiometry of the compound, the calculated value of S is $\sim$ -150 $\mu$V/K at 300 K, which gives quite good match with our experimental observed value of $\sim$ -130 $\mu$V/K. At $\mu$ $\approx$ 20 meV, we have calculated the temperature dependence of S, which is shown in fig. 6 along with the experimental data. From the figure, it is clear that in higher temperature region, a good amount of deviation ($\sim$ 50 $\mu$V/K at 620 K) between experimental and calculated values of S was observed. Now, the question arises, is there any way of improving this mBJ results. 

\begin{figure}
\includegraphics[width=0.93\linewidth, height=7.1cm]{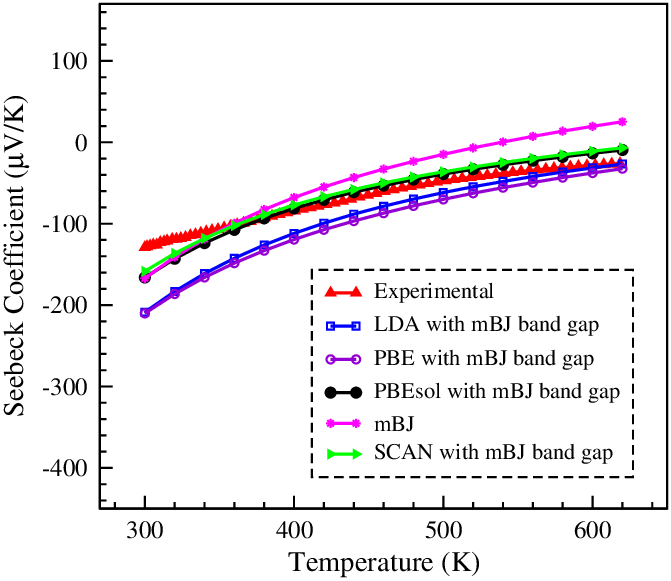} 
\caption{\small{Comparison of experimental and calculated values of S as a function of temperature.}}
\end{figure}

The main reason for failure the results of LDA, PBE, PBEsol and SCAN is they are not able to give the correct band gap for the compound, but generally they give the correct features of the band-structure. Therefore, it would be interesting to see the scenario by taking the mBJ band gap with the band-structure of other four functionals, whether the result is improving or not. This calculation is done at $\mu$ $\approx$ 20 meV, which is shown in fig. 6. From the figure, it is clear that in higher temperature region LDA and PBE give quite good match with experiment, whereas in lower temperature region calculated values are very much far away ($\sim$ 80 $\mu$V/K at 300 K) from experiment. The PBEsol and SCAN are giving good match over mBJ with experiment in the temperature range studied. Hence, the best scenario is acheived when mBJ band gap is used with the band-structure of PBEsol or SCAN for calculating the T dependent S for the compound. Therefore, these two functionals with mBJ band gap can be used for searching the new TE materials.  
\\

\section{Conclusions}
In this work, the temperature dependence of Seebeck coeficient (S) for Fe$ _{2} $VAl compound was studied in the temperature region 300$-$620 K experimentally and computationally. The measured value of S is $\sim$ -130 $ \mu $V/K at 300 K, whereas it reaches $\sim$ -26 $ \mu $V/K at 620 K. The magnitude of S decreases, as the temperature increases. The system has shown the negative value of S through the temperature range studied, which indicates the n-type behaviour of the compound. The ground state electronic calculations were performed by using five exchange-correlation (XC) functionals within density functional theory. The calculation showed, only mBJ gives the nearly accurate band gap of $\sim$ 0.22 eV, whereas all the rest four functionals predict the semi-metal like behaviour of the sample. The temperature dependence of S was also calculated by using five XC functionals and compared with the experimentally observed value. The best matching between calculated and experimental value of S vs T is observed, when mBJ band gap is used with the band-structure of the PBEsol or SCAN. Therefore, the mBJ band gap with PBEsol or SCAN band-structure can be used for searching the new high ZT thermoelctric materials.
\\
\section{Acknowledgement} 

S. Singh acknowledges support from Science and Engineering Research Board of India for financial support through the Ramanujan Fellowship and Early Career Research Award.


\begin{thebibliography}{99}

\bibitem{solarcell}
M. Z. Liu, M. B. Johnston, H. J. Snaith, Nature 2013, 501, 395

\bibitem{nuclearpower}
G. H. Marcus, Prog. Nucl. Energy 2000, 37, 5.

\bibitem{teg1}
DiSalvo F J 1999 Science 238 703

\bibitem{teg2}
Bell L E 2008 Science 321 1457

\bibitem{fossilfuel}
L. E. Doman, V. Arora, International Energy Outlook 2016, U.S. Energy Information Administration 2016.

\bibitem{progress}
C. Gayner, K. K. Kar, Prog. Mater. Sci. 2016, 83, 330

\bibitem{zt1}
Pei Y, Shi X, LaLonde A, Wang H, Chen L and Snyder G J 2011 Nature 473 66

\bibitem{zt2}
LaLonde A D, Pei Y, Wang H and Snyder G J 2011 Mater. Today 14 526


\bibitem{highzt}
Nolas G S, Sharp J and Goldsmid H J 2001 Thermoelectrics, Basic Principles and New Materials Developments (Berlin: Springer)

\bibitem{coupled}
Z. G. Chen, G. Han, L. Yang, L. Cheng, J. Zou, Prog. Nat. Sci. 2012, 22, 535.

\bibitem{bite}
O Yamashita, S Tomiyoshi, K. Makita, "Bismuth telluride compounds with high thermoelectric figures of merit", Applied Physics, vol. 93, pp. 368, 2003.

\bibitem{pbte}
K. Biswas, J. Q. He, I. D. Blum, C. I. Wu, T. P. Hogan, D. N. Seidman, V. P. Dravid, M. G. Kanatzidis, Nature 2012, 489, 414.

\bibitem{sige}
G Joshi, H Lee, Y Lan et al., "Enhanced thermoelectric figure-of-merit in nanostructured p-type silicon germanium bulk alloys", Nano Letters, vol. 8, no. 12, pp. 4670-4674, 2008.

\bibitem{kohn}
W. Kohn, L.J. Sham, Phys. Rev. 140 (1965) A1133–A1138.

\bibitem{lda92}
J.~P. Perdew and Y.~Wang {\em Phys. Rev. B}, vol.~45, p.~13244, 1992.

\bibitem{pbe}
J.~P. Perdew, K.~Burke, and M.~Ernzerhof {\em Phys. Rev. Lett.}, vol.~77,
  pp.~3865--3868, 1996.

\bibitem{pbesol}
J.~P. Perdew, A.~Ruzsinszky, G.~I. Csonka, O.~A. Vydrov, G.~E. Scuseria, L.~A.
  Constantin, X.~Zhou, and K.~Burke {\em Phys. Rev. Lett.}, vol.~100,
  p.~136406, 2008.

\bibitem{mbj}  
F. Tran, P. Blaha, Phys. Rev. Lett. 102 (2009) 226401.

\bibitem{scan}
J.~Sun, A.~Ruzsinszky, and J.~P. Perdew {\em Phys. Rev. Lett.}, vol.~115,
  p.~036402, 2015.

\bibitem{shastri} 
S.S. Shastri, S.K. Pandey, Computational Materials Science, 143 (2018), 316–324.

\bibitem{tran}
F. Tran, P. Blaha, Phys. Rev. Lett. 102 (2009) 226401.  

\bibitem{sharma}
S. Sharma, S.K. Pandey, J. Phys. D: Appl. Phys. 47 (2014) 445303

\bibitem{ddo}
D. Do \textit{et al}, PHYSICAL REVIEW B 84, 125104 (2011)

\bibitem{rai}
D P Ray \textit{et al} AIP ADVANCES 7, 045118 (2017)

\bibitem{ashcroft}
N.~Ashcroft and N.~Mermin, {\em Solid State Physics}.
\newblock Cengage Learning, 2011.

\bibitem{boscow81} K.H.J. Buschow , P.G. Van Engen 
~Journal of Magnetism and Magnetic Materials 25, 90 (1981)

\bibitem{ashutosh}
A. Patel, S.K. Pandey, Instrumentation Science and Technology, 2017, VOL. 45, NO. 4, 366–381.

\bibitem{kohn1}
P. Hohenberg and W. Kohn Phys. Rev. 136, B864 (1964).

\bibitem{kohn2}
W. Kohn and L.J. Sham, Phys. Rev. 140, A1133 (1965).

\bibitem{wein2k}
P. Blaha, K. Schwarz, G.K.H. Madsen, D. Kvasnicka, and J. Luitz, WIEN2k An Augmented Plane Wave
plus Local Orbitals Program for Calculating Crystal Properties, Karlheinz Schwarz Technische Universit at
Wien, Austria, 2001. ISBN 3-9501031-1-2.

\bibitem{boltztrap}  
G.K.H. Madsen and D.J. Singh, BoltzTraP. A code for calculating band-structure dependent quantities,
Comput. Phys. Commun. 175 (2006), pp. 67–71.

\bibitem{kurosaki}
K. Kurosaki et al. Journal of Alloys and Compounds 486 (2009) 507–510

\bibitem{positives1}
C. S. Lue and Y.-K. Kuo, Phys. Rev. B 66, 085121 (2002)

\bibitem{positives2}
M. Vasundhara, V. Srinivas, and V. V. Rao, J. Phys.: Condens. Matter 17 (2005) 6025–6036

\bibitem{positives3}
C. S. Lue, C. F. Chen, and J. Y. Lin, Phys. Rev. B 75, 064204 (2007)

\bibitem{positives4}
Y. Nishino, S. Deguchi, and U. Mizutani, Phys. Rev. B 74, 115115 (2006).

\bibitem{mikami}
M. Mikami \textit{et al.} J. Appl. Phys. 111, 093710 (2012)

\bibitem{okamura}
H. Okamura et al., Phys. Rev. Lett. 84, 3674 (2000).


\end{thebibliography}

\end{document}